\documentclass[12pt]{article}
\usepackage{amsfonts}
\usepackage{amsmath}
\usepackage{amsbsy}
\usepackage{amssymb}
\input{epsf}

\begin{document}
\voffset-10ex
\title{\bf  Influence~of~electromagnetic~fields~on~the\\
evolution~of~initially~homogeneous\\ and isotropic~universe\\[-0.5ex]}
\author{G.A.~Alekseev\footnote{e-mail: G.A.Alekseev@mi.ras.ru} \\[1ex]
{\it Steklov Mathematical Institute of Russian Academy of Sciences,} \\
{\it Gubkina st. 8, 119991, Moscow, Russia}
}
\date{}
\maketitle
\vskip-3ex

\begin{abstract}
Simple exact solutions presented here describe the universes which spatial geometries are asymptotically homogeneous and isotropic near the initial singularity, but which evolution goes under the influence of primordial magnetic fields.
In all these ``deformed'' Friedmann models (spatially flat, open or closed), the initial magnetic fields are concentrated near some axis of symmetry and their lines  are the circles  -- the lines of the azimuthal coordinate $\varphi$. Caused by the expansion of the universe, the time-dependence of a magnetic field induces (in accordance with the Faraday law) the emergence of source-free electric fields. In comparison with the Friedmann models, the cosmological expansion goes with acceleration in spatial directions across the magnetic field, and with deceleration along the magnetic lines, so that in flat and open models, in fluid comoving coordinates, the lengths of $\varphi$-circles of large enough radius or for late enough times decrease and vanish for $t\to\infty$. This means that in flat and open models, we have a partial dynamical closure of space-time at large distances from the axis, i.e. from the regions where the electromagnetic fields in our solutions are concentrated.
To get simple exact solutions of the Einstein-Maxwell and perfect fluid equations, we assume for the perfect fluid (which supports the isotropic and homogeneous ``background'' Friedmann geome\-tries) rather exotic, stiff matter equation of state $\varepsilon=p$. However, it seems reasonable to expect that  similar effects might take place in the mutual dynamics of geometry and strong electro\-magnetic fields in the universes with more realistic matter equations of state.
\end{abstract}


\subsection*{Introduction}
Our physical intuition often occurs very helpful in solving many problems of mechanics as well as theoretical and mathematical physics which can be considered in the framework of Newtonian mechanics or relativistic mechanics and electro\-dynamics in Minkowski space-time. However, it is much more difficult to adapt our intuition to the cases in which various physical phenomena take place in curved space-times, in strong gravitational, electromagnetic and other matter fields, when the behaviour of these fields is governed by more complicate nonlinear equations of Einstein's General Relativity, which implies that the  space-time geometry together with matter fields takes part in their mutual dynamics. In these cases, a great help for better understanding of this context in general can arise from consideration of some very simple exact solutions and corresponding idealized models which can be analyzed in detail providing us with useful information about qualitative behavior of gravitational, electromagnetic and other matter fields and their interactions in arbitrarily strong field regime.

In this paper, we constructed some very simple models describing the evolution of the universes under the influence of electromagnetic fields. There is a number of cosmological solutions with \textit{homogeneous} time-dependent magnetic fields, which can be found in the literature of different years, e.g. in \cite{Doroshkevich:1965}--\cite{De:1975}. Rather detailed survey of exact solutions for inhomogeneous cosmological models, including some models with magnetic fields was given later by Krasinski \cite{Krasinski:1997}.

The main features of the solutions, presented below, are that their spatial geometries at the very beginning of the cosmological expansion  \textit{asymptotically} are homogeneous and isotropic, but these include some primordial \textit{vortex} magnetic fields which are regular everywhere and which lines of force are the circles. In these solutions, the influence of electromagnetic fields on the geometry and its dynamics is negligible near the cosmological singularity, however, at later stages of the evolution, this influence becomes more ``observable'' and causes a development of large scale inhomogeneities. To simplify our models and to be able to construct the exact solutions for a complete system of the Einstein - Maxwell and perfect fluid equations, we have assumed for the perfect fluid which supports the isotropic and homogeneous ``background'' Friedmann geometry, rather exotic stiff matter equation of state $\varepsilon=p$. However, it seems reasonable to expect, that the similar effects can take place in the mutual dynamics of geometry and electro\-magnetic fields in the universes with more realistic matter equations of state.

For a construction of solutions which we are interested in, we refer first of all, to the observation of Belinski \cite{Belinski:1979} who showed that for the fields with two commuting isometries, the presence of a stiff perfect fluid does not change the dynamical part of Einstein's field equations (the projections of Einstein equations on the Killing vectors) which coincide completely with dynamical equations for vacuum. The perfect fluid parameters enter only the remaining part of the Einstein equations (``constraint equations''). Hence, for construction of solutions for this system, various methods can be used which were developed earlier for solving pure vacuum Einstein equations with two-dimensional symmetries. First of all, we mention here the inverse scattering approach and soliton generating technique of Belinski and Zakharov \cite{Belinski-Zakharov:1978} and its application in the present context in \cite{Belinski:1979}. It is  easy to see, that the same is true in the space-times with two commuting isometries, where besides gravity a stiff fluid and electromagnetic fields are present. Therefore, the symmetry reduced Einstein-Maxwell and stiff perfect fluid equations  obviously are also integrable and for solving this system we can apply the corresponding inverse scattering methods and the soliton generating technique as well as the integral equation methods developed long ago for constructing electrovacuum solutions \cite{Alekseev:1980} - \cite{Alekseev:1988} (a short survey of different methods can be found in \cite{Alekseev:2011}). However, we follow here a different way (the motivation is given in the last section of this paper).

In this paper, we consider the system of Einstein - Maxwell and stiff perfect fluid equations for the class of fields which possess only one space-like Killing vector field, i.e. for the field configurations which metric components, electromagnetic vector potential, energy density and 4-velocity of the fluid depend on three of the four space-time coordinates -- the time and two spatial coordinates. The dynamical part of these equations also coincides with the dynamical equations for electrovacuum Einstein - Maxwell fields with one Killing vector. Therefore, we can use for Einstein-Maxwell and stiff perfect fluid equations the solution generating methods (based on the internal or ``hidden'' symmetries of these equations) which were used earlier for  electrovacuum Einstein - Maxwell equations. Though this last mentioned system has not been found to be integrable, it is known that it can be presented in the form of generalized Ernst equations for two complex Ernst-like potentials. These equations admit some symmetry transformations which constitute the 8-parametric $SU(2,1)$ group \cite{Neugebauer-Kramer:1969} - \cite{Kinnersley:1973}. Therefore, these are also the symmetry transformations of the system of Einstein - Maxwell and stiff perfect fluid equations with one space-like Killing vector field.
For our present purposes we use a particular type of electrovacuum $SU(2,1)$-symmetry transformations which were discovered earlier by Harrison \cite{Harrison:1968}. These transformations allow to generate from the solutions for pure gravity (and with stiff perfect fluid solutions, in our case) the corresponding solutions which include also some electromagnetic fields.

In the subsequent sections of this paper, we describe the three-dimensional Ernst-like equations and the Harrison symmetry transformations for Einstein - Maxwell and stiff perfect fluid
equations for space-times with one Killing vector field. Also, we describe in detail the properties of the solutions which arise as the result of application of Harrison transformations to different types of the homogeneous and isotropic Friedmann solutions with a stiff perfect fluid and with open, flat and closed spatial geometries. A set of very simple solutions which arise in this way, show a number of interesting phenomena which take place in the interaction of strong gravitational and electromagnetic fields and which are described in the subsequent sections of the paper. Some important remarks and comments are given in the ``Concluding remarks''.

\subsubsection*{Friedmann cosmological models with stiff matter}
The space-times with spatially isotropic and homogeneous distributions of matter (ideal fluid) are described by one of the Friedmannn models with the metric
\begin{equation}\label{Friedmann}
ds^2=a^2(t)\left[dt^2-d\chi^2-S^2(\chi)(d\theta^2 +\sin^2\!\theta d\varphi^2) \right]
\end{equation}
where the function $S(\chi)$ determines one of three possible types of the geometry of these models, in which the three-dimensional physical spaces can be opened, flat or closed spaces of (spatially) constant curvature (evaluating, however, with time). The function $a(t)$ determines the dynamics of the model in accordance with the Einstein equations, provided the matter equation of state is given.
For stiff matter fluid (the equation of state $\varepsilon=p$), the Einstein equations determine $a(t)$, $S(\chi)$ and the energy density $\varepsilon(t)$:
\begin{equation}\label{aS-functions}
\left\{\!\!\begin{array}{llcl}
a(t)=a_0 \sqrt{\sinh (2 t)},&S(\chi)=\sinh\chi,&&k=-1\\[1ex]
a(t)=a_0\,\sqrt{2t},&S(\chi)=\chi,&&k=0\\[1ex]
a(t)=a_0\sqrt{\sin(2 t)},&S(\chi)=\sin\chi,&&k=1
\end{array}\right.\hskip1ex\text{and}\quad \varepsilon=\dfrac{3 a_0^4}{8\pi a^6(t)}
\end{equation}
where $k=-1,0,1$ respectively for opened, flat and closed Friedmann models.
\[\]

\subsubsection*{The Einstein-Maxwell and stiff fluid equations}
\medskip
To construct a new solutions, we consider the Einstein - Maxwell and stiff fluid equations in the absence of any charges and currents ($i,k,\ldots=1,\ldots,4$; $\gamma=1$, $c=1$):
\begin{equation}\label{EMsfEquations}
\begin{array}{l}R_{ik}-\dfrac 12 R g_{ik}=8\pi (T_{ik}^{\scriptscriptstyle{(EM)}}+T_{ik}^{(\varepsilon=p)})\\[2ex]
\nabla_k F_i{}^k=0,\quad \nabla_{[i} F_{jk]}=0,
\end{array}
\end{equation}
where $T_{ik}^{EM}$ and $T_{ik}^{(\varepsilon=p)}$ are the energy-momentum tensors of electromagnetic field and stiff matter fluid respectively:
\[\begin{array}{l}
T_{ik}^{\scriptscriptstyle{(EM)}}=-\dfrac 1{4\pi}[F_{il} F_k{}^l-\dfrac 14 F_{lm} F^{lm} g_{ik}],\quad F_{ik}=\partial_i A_k-\partial_k A_i \\[2ex]
T_{ik}^{(\varepsilon=p)}=(\varepsilon+p)u_i u_k-p g_{ik},\quad \varepsilon =p,\quad u_k u^k=1,
\end{array}
\]
Here $\varepsilon$ and $p$ are respectively the energy density and pressure of a fluid and $u^k$ is a 4-velocity of a fluid. The equations (\ref{EMsfEquations}) can be presented also in the form
\begin{equation}\label{EMsfEqs}
R_{ik}=8\pi (T_{ik}^{\scriptscriptstyle{(EM)}}+ 2\varepsilon u_i u_k),\qquad
\nabla_k F_i{}^k=0,\quad F_{ik}=\partial_i A_k-\partial_k A_i.
\end{equation}

\subsubsection*{\small Dynamical restrictions on the fluid motion}
The dynamical equations of the fluid motion are determined by the conservation law
$\nabla_k T^{ik}=0$ and can be reduced to the equations
\begin{equation}\label{EOM}\nabla_k(\sqrt{\varepsilon} u^k)=0,\qquad
u^k\nabla_k (\sqrt{\varepsilon}u^i)=\nabla^i\sqrt{\varepsilon}
\end{equation}
We assume also that the motion is curl-free, i.e. there exists a potential $\phi$ such that
\[\sqrt{\varepsilon} u^i=\nabla^i\phi\qquad\Longrightarrow\qquad \nabla_k\nabla^k\phi=0,\quad\varepsilon=\nabla^k\phi\nabla_k\phi
\]
and in this case, the equations of motion (\ref{EOM}) are satisfied identically.

\subsubsection*{Metrics admitting a Killing vector field.}
Let us assume now that the space-time metric admits one time-like or space-like Killing vector field $\xi^i=\delta^i_4$. Then it can be parameterized as follows:
\begin{equation}\label{GeneralMetric}
\begin{array}{l}
ds^2 = \epsilon_0 H (dx^4 +\Omega_\alpha dx^\alpha)^2-
\epsilon_0 H^{-1}\, {\gamma_{\alpha\beta}}\, dx^\alpha dx^\beta,\\[2ex]
\xi^i=\delta_4{}^i,\qquad \xi_i=\epsilon_0 H \{\Omega_\alpha,1\}
\end{array}
\end{equation}
where $\alpha,\beta,\ldots=1,2,3$ and $\epsilon_0=\pm 1$. The value $\epsilon_0 H$ is the norm of the Killing vector field $\xi^i$ so that $\epsilon_0=1$ for the time-like vector and $\epsilon_0=-1$ for the space-like one; $\gamma_{\alpha\beta}$ is a (conformal) metric on  3-space orthogonal to $\xi^i$.
The metric functions $H\ge 0$, $\Omega_\alpha$ and $\gamma_{\alpha\beta}$ depend only on $x^1$, $x^2$ and $x^3$.

\subsubsection*{\small Kinematic restriction on the stiff matter configurations.}
In what follows, we assume that there is no any motion of a stiff matter fluid along the Killing vector field, i.e.
\begin{equation}\label{OrthogonalMotion}
\xi^k u_k=0,\qquad\Rightarrow\qquad \epsilon_0=-1,\quad \xi_i\xi^i=-H
\end{equation}
The last two of the equations (\ref{OrthogonalMotion}) follow from the first one and from the condition that $u^k$ is a time-like vector. In the coordinates in which the metric takes the form (\ref{GeneralMetric}), the components $u^i$ of 4-velocity of the fluid are
\[u_i=\{u_\alpha,\,0\},\qquad
u^i=\{H \gamma^{\alpha\beta} u_\beta,\,-H \gamma^{\gamma\delta} \Omega_\gamma u_\delta\}
\]
where the matrix $\Vert\gamma^{\alpha\beta}\Vert$ is inverse to $\Vert\gamma_{\alpha\beta}\Vert$ and three components $u_\alpha$ remain arbitrary. However, in the above we assumed that the fluid motion possess a potential. This implies that the potential
$\phi$ is independent on $x^4$ and the components of 4-velocity and the energy density possess the expressions
\[u_i=\varepsilon^{-1/2}\{\partial_\alpha\phi,\,0\},\quad u^i=H \varepsilon^{-1/2}\{
\gamma^{\alpha\gamma}\partial_\gamma\phi,\,-\gamma^{\gamma\delta}\Omega_\delta \partial_\delta\phi\},\quad\varepsilon=H \gamma^{\gamma\delta}\partial_\gamma\phi\, \partial_\delta\phi.
\]

\subsubsection*{\small The Ernst-like form of the symmetry reduced field equations.}
Using $h_i{}^j\equiv\delta_i{}^j-(\epsilon_0 H)^{-1} \xi_i\xi^j$ as a projector on the 3-space orthogonal to the Killing vector, it is convenient to split the equations (\ref{EMsfEqs}) into three (coupled to each other) parts corresponding to the projections of the Ricci tensor
\begin{equation}\label{RicciSplit}
R_{jl}\xi^j\xi^l,\qquad R_{jl}\xi^j h_i{}^l,\qquad R_{jl} h_i{}^j h_k{}^l
\end{equation}
In the absence of a fluid, i.e. for electrovacuum Einstein - Maxwell equations, it is known (see Neugebauer and Kramer \cite{Neugebauer-Kramer:1969}, Israel and Wilson \cite{Israel-Wilson:1972} and Kinnersely \cite{Kinnersley:1973}) that the equations corresponding to the first two projections in (\ref{RicciSplit}), together with the Maxwell equations, can be presented in the form which is a three-dimensional analogue of the well known two-dimensional Ernst equations for space-times which admit two commuting isometries. On the other hand, in the absence of electromagnetic fields, in two-dimensional case of gravity with stiff matter fluid, Belinski \cite{Belinski:1979} observed that the stiff fluid does not give any input into the dynamical part of vacuum Einstein equations if there is no any motion  of this fluid along two commuting Killing vector fields. It is easy to join these two observations and show that in three-dimensional case, in the presence of both, electromagnetic fields and stiff matter fluid, if there is no any motion of the stiff matter fluid along the Killing vector field, the field equations corresponding to the first two projections in (\ref{RicciSplit}), together with the Maxwell equations, do not include any input from the fluid and take exactly the same form as generalized Ernst equations \cite{Neugebauer-Kramer:1969} --\cite{Kinnersley:1973} for electrovacuum space-times with one Killing vector field, while the equations corresponding to the third projection in (\ref{RicciSplit}) lead to three-dimensional tensor equations for the metric $\gamma_{\alpha\beta}$. As a result, we obtain a complete system
\begin{eqnarray}\label{ErnstEquations}
\left\{\!\!\begin{array}{l}
(\text{Re}\mathcal{E}+\Phi\overline{\Phi})\gamma^{\alpha\beta}\nabla_{\alpha} \nabla_\beta\mathcal{E}= \gamma^{\alpha\beta}(\nabla_\alpha\mathcal{E}+2\overline{\Phi}\nabla_\alpha\Phi) \nabla_\beta \mathcal{E}\\[1ex]
(\text{Re}\mathcal{E}+\Phi\overline{\Phi})\gamma^{\alpha\beta}\nabla_{\alpha} \nabla_\beta\Phi= \gamma^{\alpha\beta}(\nabla_\alpha\mathcal{E}+2\overline{\Phi}\nabla_\alpha\Phi) \nabla_\beta \Phi\\[2ex]
R_{\alpha\beta}[\gamma]=\dfrac
{(\nabla_{(\alpha}\mathcal{E}+2\overline{\Phi}\nabla_{(\alpha}\Phi) (\nabla_{\beta)}\overline{\mathcal{E}}+2\Phi\nabla_{\beta)} \overline{\Phi})}{2(\text{Re}\mathcal{E}+\Phi\overline{\Phi})^2}
-\dfrac{2\nabla_{(\alpha}\Phi \nabla_{\beta)}\overline{\Phi}}{(\text{Re}\mathcal{E}+\Phi\overline{\Phi})}+16\pi \nabla_\alpha \phi \nabla_\beta \phi\\[2ex]
\gamma^{\alpha\beta}\nabla_\alpha\nabla_\beta\phi=0
\end{array}\right.
\end{eqnarray}
where a bar denotes a complex conjugation, $\nabla_\alpha$ is a covariant derivative with respect to the 3-metric $\gamma_{\alpha\beta}$ and $R_{\alpha\beta}[\gamma]$ is the Ricci tensor for this metric. In contrast to the two-dimensional case, the equations (\ref{ErnstEquations}) are coupled: the unknown metric $\gamma_{\alpha\beta}$ enters the first two  equations (\ref{ErnstEquations}), while the Ernst potentials and their first derivatives enter the ``source'' terms in the right-hand sides of the third, tensor equation.

The solution $\{\mathcal{E},\, \Phi,\,\gamma_{\alpha\beta}\}$ of the equations  (\ref{ErnstEquations}) allow to determine all components of metric, electromagnetic potential, 4-velocity of the fluid and its energy density:
\begin{equation}\label{Components}\begin{array}{l}
\left\{\begin{array}{l}
\epsilon_0 H=\text{Re}\,\mathcal{E}+\Phi\,\overline{\Phi},\\
\partial_\alpha\Omega_\beta-\partial_\beta\Omega_\alpha= -\epsilon_0 H^{-2}\varepsilon_{\alpha\beta}{}^\gamma\text{Im} \bigl[\partial_\gamma\mathcal{E}+2\overline{\Phi}\, \partial_\gamma\Phi\bigl],
\end{array}\right.\\[3ex]
\left\{\begin{array}{l}
A_4=\text{Re}\,\Phi,\\
\partial_\alpha A_\beta-\partial_\beta A_\alpha=-\Omega_\alpha\partial_\beta\text{Re}\Phi+ \Omega_\beta\partial_\alpha\text{Re}\Phi-H^{-1}\varepsilon_{\alpha\beta}{}^\gamma \partial_\gamma\text{Im}\Phi,
\end{array}\right.\\[3ex]
\quad u_i={\varepsilon}^{-1/2}\{\partial_\alpha\phi,\,0\},\qquad
\varepsilon=H\,\gamma^{\alpha\beta}\partial_\alpha\phi\, \partial_\beta\phi\,.
\end{array}
\end{equation}

\vspace{-1ex}

\subsubsection*{\small Non-gauge symmetries of the dynamical equations  (\ref{ErnstEquations})}

Similarly to the electrovacuum case, the dynamical equations (\ref{ErnstEquations}) can be written in a complex vector form. For this, following \cite{Kinnersley:1973}, we introduce instead of two complex potentials $\mathcal{E}(x^\alpha)$, $\Phi(x^\alpha)$, three unknown complex variables $u(x^\alpha)$, $v(x^\alpha)$ , $w(x^\alpha)$ so that
\[\mathcal{E}=\dfrac{u-w}{u+w},\qquad \Phi=\dfrac{v}{u+w}.
\]
Keeping in mind that we have a freedom to impose one more constraint on these three new variables, we can write, in complete analogy with \cite{Kinnersley:1973} the equations (\ref{ErnstEquations}) in terms of a complex vector function $Y^A=\{u,\,v,\,w\}$  ($A,B,\ldots=1,2,3$) in the form
\begin{eqnarray}\label{YEquations}
\left\{\!\!\begin{array}{l}
(\overline{Y}_C\,Y^C)\,\gamma^{\alpha\beta}\nabla_\alpha\nabla_\beta Y^A-2 \gamma^{\alpha\beta}(\overline{Y}_C\nabla_\alpha Y^C)\,\nabla_\beta Y^A=0\\[2ex]
R_{\alpha\beta}[\gamma]=\dfrac{2\overline{Y}_A Y_B\,\partial_{(\alpha} Y^A\partial_{\beta )} \overline{Y}^B-2(\overline{Y}_A\,Y^A)\,\partial_{(\alpha} Y_B\,\partial_{\beta )} \overline{Y}^B}{(\overline{Y}_C\,Y^C)^2}
+16\pi \phi_{,\alpha} \phi_{,\beta}\\[2ex]
\gamma^{\alpha\beta}\nabla_\alpha\nabla_\beta\phi=0
\end{array}\right.
\end{eqnarray}
where $Y_A =\eta_{AB} Y^B$ with $\eta_{AB}=\text{diag}\{1,\,1,\, -1\}$.

A beautiful property of the equations (\ref{YEquations}) is that these are invariant under the linear transformation of the unknowns, provided it  leaves the ``metric'' $\Vert\eta\Vert$ invariant:
\[Y^A\to U^A{}_B Y^B,\quad\gamma_{AB}\to\gamma_{AB},\quad \phi\to \phi\qquad
\left.\vphantom{Y_A^B}\right\Vert\qquad  U^C{}_A\,\eta_{CD}\,\overline{U}^D{}_B=\eta_{AB}
\]
where $\mathbf{U}=\Vert U^A{}_B\Vert$ is a constant complex matrix. The last of the equations just above means that the matrices of these transformations constitute a group which can be identified with $SU(2,1)$. It is known, that the elements of this group can be parameterized by eight independent real parameters \cite{Kinnersley:1973}. Some of these parameters are pure gauge ones, i.e. the corresponding transformations do not change the space-time geometry as well as the properties of the matter fields, while the others can transform one solution to a physically different one. In particular, such transformations can transform static solution to a stationary one, the solution for waves with linear polarization to a solution which includes the waves with different polarizations (the Ehlers transformations) or pure vacuum solutions to solutions with electromagnetic fields (Harrison transformations \cite{Harrison:1968}). The last one will be the most  interesting for us farther.

\subsubsection*{Harrison transformation}
Following \cite{Kinnersley:1973}, the Harrison transformation can be presented in the form
\begin{equation}\label{Harrison}\left.\begin{array}{l}
(u+w)\to (u+w)-2\overline{c}\, v-c\overline{c}(u-w)\\[1ex]
v\to v+c (u-w)\\[1ex]
(u-w)\to(u-w)
\end{array}\hskip1ex\right\Vert\hskip1ex
U^A{}_B=\begin{pmatrix}
1-\dfrac {c\overline{c}}2 &-\overline{c}&\dfrac {c\overline{c}}2\\
c&1&-c\\
-\dfrac {c\overline{c}}2&-\overline{c}&1+\dfrac {c\overline{c}}2
\end{pmatrix}
\end{equation}
This implies the following transformation of the Ernst potentials
(everywhere below ``$\circ$'' denotes the values of potentials of the transforming solution):
\[
\mathcal{E}=\dfrac{\overset \circ{\mathcal{E}}}{1-2\overline{c}\, \overset \circ{\Phi}-c\overline{c}\,\overset \circ{\mathcal{E}}},\qquad
\Phi=\dfrac{\overset \circ{\Phi}+c\,\overset \circ{\mathcal{E}}}{1-2\overline{c}\, \overset \circ{\Phi}-c\overline{c}\,\overset \circ{\mathcal{E}}}
\]
The corresponding transformation of $H$ and the energy density take the form
\[H=\dfrac{\overset \circ{H}}{\vert{1-2\overline{c}\, \overset \circ{\Phi}-c\overline{c}\,\overset \circ{\mathcal{E}}}\vert^2},\qquad
\varepsilon=\dfrac{\overset \circ{\varepsilon}}{\vert{1-2\overline{c}\, \overset \circ{\Phi}-c\overline{c}\,\overset \circ{\mathcal{E}}}\vert^2},
\]
However, the transformations of the metric functions $\Omega_\alpha$ and the components of the Maxwell tensor for electromagnetic field do not possess an algebraic character and these can be determined from general expressions (\ref{Components}).

\subsubsection*{Friedmann universes in terms of the Ernst potentials}
Our purpose now is to apply the Harrison transformation described just above to the Friedmann solutions (\ref{Friedmann})-(\ref{aS-functions}). For this, we determine at first, the metric functions, Ernst potentials and matter fields for solutions (\ref{Friedmann}).
For these solutions we choose $\{x^1,x^2,x^3,x^4\}=\{t,\chi,\theta,\phi\}$ and therefore,
the Ernst potentials and the metric on the 3-dimensional sections orthogonal to the Killing vectors are
\[\begin{array}{l}
\overset\circ{\mathcal{E}}=-a^2(t) S^2(\chi)\sin^2\theta,\\[1ex]
\overset\circ{\Phi}=0,
\end{array}\quad
\gamma_{\alpha\beta}=a^4(t) S^2(\chi)\sin^2\theta\begin{pmatrix}1&0&0\\0&-1&0\\
0&0&-S^2(\chi)\end{pmatrix}.
\]
Besides that, for other metric functions and a fluid potential we have
\[\overset\circ{H}=a^2(t) S^2(\chi)\sin^2\theta,
\quad\overset\circ{\Omega}_\alpha=0,\quad \phi=\sqrt{\frac{3}{32\pi}}\dfrac{a_0^2}{a^2(t)},\]
where the functions $a(t)$ and $S(\chi)$ were described in (\ref{aS-functions}).

\subsubsection*{Friedmann universes deformed by electromagnetic fields}
The Harrison transformation (\ref{Harrison}) of the Friedmann solutions (\ref{Friedmann})-(\ref{aS-functions}) leads to the solutions for Friedmann universes which metrics are deformed by electromagnetic fields:
\begin{equation}\label{Metric}
ds^2=a^2(t)\,\Lambda^2\Bigl[dt^2-d\chi^2-S^2(\chi)d\theta^2\Bigr]-
\dfrac{a^2(t)}{\Lambda^2}S^2(\chi)\sin^2\!\theta d\varphi^2.
\end{equation}
The electromagnetic fields in these solutions are described by the real vector potential $\mathbf{A}$ or, equivalently, by a complex scalar potential $\Phi$ of the form
\begin{equation}\label{EHfields}
\mathbf{A}=-2 H_0 a^2(t) S^2(\chi)\cos\theta\,\{\dfrac{S^\prime(\chi)}{S(\chi)},\,
\dfrac{\dot a (t)}{a(t)},\,0,\,0\,\},\quad \Phi=\dfrac{i H_0 a^2(t) S^2(\chi)\sin^2\theta} {\Lambda}
\end{equation}
and the pressere and the energy density of the fluid possess the expressions
\begin{equation}\label{pe}
p=\varepsilon=\dfrac{3 a_0^4}{8\pi a^6(t) \Lambda^2},\qquad\text{where}\qquad
\Lambda=1+H_0^2 a^2(t) S^2(\chi)\sin^2\theta.
\end{equation}
Here the parameter $c$ of Harrison transformation was chosen pure imaginary $c=-i H_0$. Another its choice (with $c\overline{c}=H_0^2$) would not change the metric (\ref{Metric}) but it would lead to a dual transformation of electromagnetic fields such that primordial electric field also appears in the solutions.

\subsubsection*{Properties of Friedmann universes with electromagnetic fields}
In the solutions (\ref{Metric}), the metric and electromagnetic fields possess the axial symmetry and depend on time and two spatial coordinates (concerning other symmetries of these solutions see the ``Concluding remarks'' section).

{\it The initial singularity.} The metric (\ref{Metric}) possess an initial singularity at $t=0$ which is similar to that in Friedmann solutions (\ref{Friedmann}). Near this singularity, i.e. for $a(t)\to 0$, we have $\Lambda\to 1$ and the metric (\ref{Metric}) asymptotically becomes homogeneous and isotropic, coinciding with Friedmann metric (\ref{Friedmann}).

{\it Electromagnetic fields measured by a local observer.} In our solution, the components of the Maxwell tensor possess the expressions:
\[F^{ik}=-\dfrac{2 H_0}{a^2(t)\Lambda^4}\begin{pmatrix}
0&\cos\theta&-\dfrac{S^\prime(\chi)}{S(\chi)}\sin\theta&0\\[1ex]
-\cos\theta&0&\dfrac{\dot a(t)}{a(t)}\sin\theta&0\\
\dfrac{S^\prime(\chi)}{S(\chi)}\sin\theta&-\dfrac{\dot a(t)}{a(t)}\sin\theta&0&0\\
0&0&0&0
\end{pmatrix}
\]
Using the metric (\ref{Metric}), we introduce the orthonormal basis of one-forms
\[
e{}_{\widehat{t}}=\Lambda a(t) dt,\hskip1.5ex
e{}_{\widehat{\chi}}=\Lambda a(t) d\chi,\hskip1.5ex
e{}_{\widehat{\theta}}=\Lambda a(t) S(\chi) d\theta,\hskip1.5ex
e{}_{\widehat{\varphi}}=\dfrac{a(t)}{\Lambda}S(\chi)\sin\theta d\varphi\]
The projections of Maxwell tensor on this basis determine the stresses of magnetic and electric fields measured by a local fluid comoving observer:
\begin{equation}\label{LocalEH}
\begin{array}{lccl}
H{}^{\widehat{\chi}}=0,&&&E{}^{\widehat{\chi}}=\dfrac{2 H_0\cos\theta}{\Lambda^2},\\[1.5ex]
H{}^{\widehat{\theta}}=0,&&&E{}^{\widehat{\theta}}=-\dfrac{2 H_0 S^\prime(\chi)\sin\theta}{\Lambda^2},\\[1ex]
H{}^{\widehat{\varphi}}=\dfrac{2 H_0 \dot{a}(t) S(\chi)\sin\theta}{a(t)\Lambda^2},&&&E{}^{\widehat{\varphi}}=0.
\end{array}
\end{equation}

{\it The structure of primordial magnetic field.}
From these expressions, one can see that in all models the magnetic field is vortex, it is completely regular everywhere and its lines coincide with the closed coordinate lines of the azimuthal coordinate $\varphi$.
The magnitude of this magnetic field vanishes on the axis of symmetry $\theta=0$ and $\theta=\pi$, then it grows with the distant from this axis, reaches a maximum and for larger distances it decreases and vanishes at another pole of 3-sphere in closed model or vanishes asymptotically at spatial infinity in flat and open models. The spatial structure of the magnetic field on a surface $t=const$ is shown on the Figure 1.

\begin{figure}[]
\begin{center}
\epsfxsize=0.75\textwidth \epsfbox{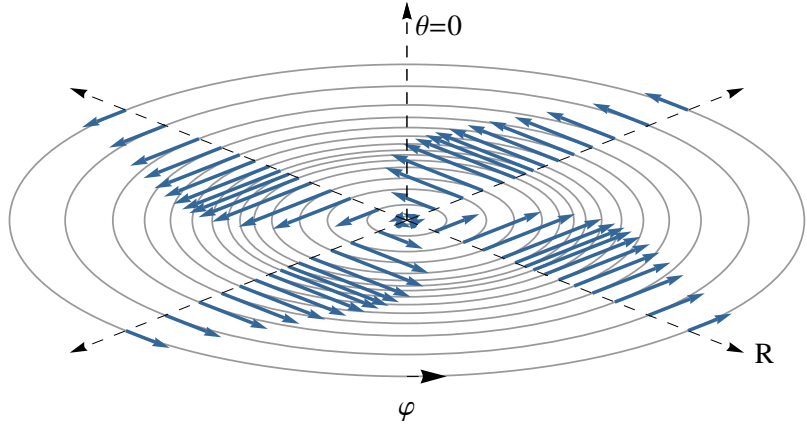}
\end{center}
\caption{\footnotesize The axisymmetric structure of the magnetic field in the solutions (\ref{Metric}) -- (\ref{LocalEH}) is shown here on the sections orthogonal to the axis $(\theta=0,\pi)$ on the space-like hypersurfaces $t=const$. The arrows show the 3-vectors of the magnetic field which possess only one non-vanishing component
$H^{\widehat{\varphi}}=\left(\dfrac{2\dot a}{a^2}\right) \dfrac{R}{(1+R^2)^2}$, where $R=H_0 a(t) S(\chi)\sin\theta$.
The lines of force of this magnetic field are the lines of azimuthal coordinate $\varphi$. The length of arrows characterizes the magnitude of magnetic field.}
\end{figure}

{\it The structure of induced electric field.} In accordance with the Faraday law, the time dependence of the magnetic field (caused by the cosmological expansion) induces the emergence of a source-free and regular everywhere electric field which provides the electric field flux through the magnetic field ($\varphi$-line) contours.
The non-vanishing components of the electric field are $E^\chi$ and $E^\theta$ and therefore, the spatial 3-vectors of electric field are tangent to the two-dimensional surfaces $\varphi=const$. To understand the structure of this field, we should find the lines of this vector field on these surfaces. The equation for electric lines of force and its solution take the forms
\[\dfrac{d\theta}{d\chi}=\dfrac{F^{\theta t}}{F^{\chi t}}=-\dfrac{S^\prime(\chi)}{S(\chi)}
\tan \theta \quad\Rightarrow\quad S(\chi)\sin\theta=const.
\]
This solution shows that in the case of Friedman model with a flat spatial geometry the electric field lines are the straight lines which are orthogonal to the circle $\varphi$-lines and parallel to the axis $\theta=0$ and $\theta=\pi$. In the case of open models, the topology of the lines of electric field is the same, while the case of closed model is more interesting. In this case, similarly to the lines of magnetic field, the lines of electric field are also closed and they go along the parallels on the quasi-spherical surfaces with coordinates $(\chi,\theta)$,  however, these parallels surround the axis which is orthogonal to the axis $\theta=0$ and $\theta=\pi$ and to the magnetic field lines.
These configurations of electromagnetic fields are concentrated mainly in the cylindrical region near the axis of symmetry, resemble very much the coil which turns represent the circle lines of the magnetic field and the electric lines go inside of this coil parallel to its axis.

{\it The electromagnetic input into the energy density.}
The expressions given above show also that near the singularity
the main input to the electromagnetic energy density is given by the only non-vanishing component of the magnetic field which goes to infinity as $1/t$, while the electric field input is much less because its components remain finite when $t\to 0$. This allows us to say that the role of primordial field in our solution is played by the magnetic field and the electric field arises due to Faraday induction. On the other hand, even the infinitely growing magnetic field makes the input into the energy density which is much less than the energy density of the fluid: the former grows as $t^{-3}$ for $t\to 0$, while the value of the magnetic field input into the energy density is of the order $t^{-2}$. This allows the universe to be initially (for small $t$) homogeneous and isotropic due to  homogeneous and isotropic initial distribution of stiff fluid.

{\it Behaviour of the space-time geometries in spatial directions.}
On the axis of symmetry ($\theta=0,\pi$) the metrics (\ref{Metric}) at any time coincides with the corresponding metrics of Friedmann solutions without electromagnetic fields. An important deviation of the geometry (\ref{Metric}) from Friedmann one takes place in flat and open models at spatial infinity in the directions $S(\chi)\sin\theta\to\infty$. In the regions far from the axis $\theta=0,\pi$ in the open models, deviations from open Friedmann universes possess crucial character: the spatial geometry asymptotically become closed in these directions. This closure of spatial metric become evident because of the non-monotonous behaviour of the lengths of the circles which are the coordinate lines of $\varphi$ with growing their ``radius'' $\chi$ (the length of these lines are determined by the value of  $\sqrt{-g_{\varphi\varphi}}$  in (\ref{Metric})). Indeed, for small enough values of $\chi$ (or, equivalently, of $S(\chi)$) the lengths of these lines increase if $\chi$ increases. However, for farther increasing of $\chi$ (with non-vanishing $\sin\theta$) the value of $\Lambda$ becomes proportional to $S(\chi)^2\sin^2\theta$ and therefore,  $g_{\varphi\varphi}$ is proportional there to $S(\chi)^{-2}\sin^{-2}\theta$ and  vanishes asymptotically at spatial infinity at these directions.

{\it Spatial closure of Melvin and Bertotti-Robinson static magnetic universes.}
Partial spatial closure of the geometry (\ref{Metric}) described above resembles very much the spatial closure of a static Melvin magnetic universe in spatial directions orthogonal to its axis of symmetry. In Melvin universe, even though the value of magnetic field (parallel to the axis of symmetry and constant along this axis) decrease with the distance from this axis, the anisotropic electromagnetic stresses in this ``bundle of magnetic lines'' are so strong that these cause the closure of the spatial geometry in the directions orthogonal to the symmetry axis. Another example of such closure take place in the static Bertotti-Robinson electromagnetic universe. The physical space of this universe is filled by completely homogeneous (spatially constant) magnetic/electric field. In this case, the closure of spatial geometry takes place even not at spatial infinity, but at finite distance from the axis of symmetry. As a result, this model possess $AdS^2\times S^2$ space-time geometry (where $AdS^2$ means two-dimensional Anti-de Sitter space-time), so that the internal geometry on the sections $(t=const,\,z=const)$ is homogeneous and isometric to the metrics on the 2-spheres of constant radius which is inverse proportional to the stress of a magnetic/electric field.

{\it Acceleration of cosmological expansion and dynamical closure of flat and open models.}
As it was mentioned above, on the axis of symmetry ($\theta=0,\pi$) the metrics (\ref{Metric}) at any time $t$ coincides with the corresponding metrics of Friedmann solutions. Due to the presence in all metric components of the factor $\Lambda^2$, besides $g_{\varphi\varphi}$, the cosmological expansion outside the axis $\theta\ne 0,\pi$ goes \emph{faster} than in the corresponding Friedmann models in all spatial directions \emph{across} the magnetic field.

An interesting phenomena can be seen in these space-times, if we consider the character of the cosmological expansion along the magnetic line circles.
In contrast with the directions across the magnetic field, the expansion \emph{along} the magnetic field lines goes more slowly than in the Friedmann models due to the presence of the factor $\Lambda^{-2}$ in $g_{\varphi\varphi}$. At earlier stage of the cosmological expansion, the lengths of magnetic line circles, considered as functions of time for finite comoving ``radius'' $\chi$, grows with time. However, later even in flat and open models, the decelerated expansion along the magnetic field lines ($\varphi$-lines with $\chi=const$) changes to a contraction: the lengths of $\varphi$-circles with given comoving ``radius'' $\chi$ pass through a maximum and then begins to decrease with time and vanishes asymptotically for $t\to\infty$, while the cosmological expansion in other spatial directions in flat and open models continues until $t=\infty$ and the actual ``radius'' of these circles will continue to grow with acceleration.
This means that under the influence of electromagnetic fields (\ref{LocalEH}) on the dynamics of these universes, their space-times suffer partial dynamical closure.

Speculatively, it seems interesting to note that in our models the acceleration of the cosmological expansion take place in almost all radial directions from the observer, located somewhere in the region where the electromagnetic fields are concentrated, and it can be measured by some ``ordinary'' methods, while the behaviour in time of the lengths of the circles of very large radius can be much more difficult to be measured and this can happen to be out of the observers' attention.

\subsubsection*{Concluding remarks}
(1) The metrics (\ref{Friedmann}) depend on three coordinates $t$, $\chi$ and $\theta$ and obviously admit the Killing vector field $\xi_{\varphi}=\partial/\partial_\varphi$. Just with this Killing vector field we have associated the Ernst potentials and their Harrison transformation used in our construction. However, the Friedmann metrics  admit other isometries and in particular, another one, say $\xi_{\psi}$ commuting with $\xi_{\varphi}$. For a benefit of the readers, we recall here the known transformation of coordinates (see, for example, \cite{Belinski:1979}) which leads to the forms of Friedmann universes which depend on two coordinates only. Namely, the coordinate transformation $\{\chi,\theta\}\to \{x,\psi\}$ determined by the expressions
\[\left\{\begin{array}{l}
S(\chi)\sin\theta=X(x),\\[1ex]
S(\chi)\cos\theta=X^\prime(x) Y(\psi),
\end{array}\right.\quad
\{X(x),Y(\psi)\}=\left\{\begin{array}{ll}
\{\sinh x,\sinh \psi\},&k=-1,\\
\{x,\psi\},&k=0,\\
\{\sin x,\sin \psi\},&k=1.
\end{array}\right.
\]
leads from the metrics (\ref{Friedmann}) to the metrics which depend only on $t$ and $x$

\[ds^2=a^2(t)\left[dt^2-dx^2-X^2(x)\,\, d\varphi^2-{X^\prime}^2(x)\,\, d\psi^2\,\right]
\]
and obviously admit two commuting isometries $\xi_{\varphi}=\partial/\partial_\varphi$ and $\xi_{\psi}=\partial/\partial_\psi$.
Using the formalism described above, we can associate the Ernst potentials and their Harrison transformation not with $\xi_{\varphi}$ only (as it was done in this paper) but with arbitrary linear combination of these two Killing vector fields. This will lead to the solutions which describe the Friedmann universes deformed by electromagnetic fields of more complicate
and, probably, more rich and interesting structures.

(2) As we have seen in this paper, even a simple scalar field (which was alternatively interpreted here as the potential for the stiff fluid motion) can play an important role in the evolution of the universe. In the modern literature, various gravity models (e.g., string gravity and supergravity models) are considered in four and higher dimensions. The symmetry reductions of these models include a large number of ``moduli fields'' - the scalar fields of different structures and couplings as well as various gauge fields. These (symmetry reduced) models admit a large number of ``hidden'' symmetries. Many of these symmetries (including the Harrison-type transformations) were used by different authors for constructing various charged black hole solutions in these models (``charging transformations''). Our present construction shows that similar transformations can give rise to simple exact solutions which describe the evolution of various cosmological models in these theories under the influence of various (spatially non-singular) scalar and gauge fields.

(3) To conclude our discussion of the solutions constructed in this paper, it is necessary to mention that it is difficult to assign to these solutions immediate astrophysical interpretation as relevant to our Universe. This is caused mainly by two important circumstances: the ``exotic'' character of used equation of state $p=\varepsilon$ and an existence of tremendous observational restrictions on the possible anisotropy of our Universe. Concerning the first one, as it was already mentioned in the introduction, this equation of state was used for pure technical reason - to get the field equations which possess the ``hidden'' symmetries including the Harrison transformations, but in the presence of the ideal fluid which ``supports'' Friedmann cosmological background. However, one may expect that the presence of electromagnetic fields in the universes with more realistic matter equations of state may give rise to similar physical phenomena caused by nonlinear interaction of gravitational and electromagnetic fields, which we can study in our solutions in many details.

What about the observationally restricted possibility for our universe to be anisotropic due to existence of primordial magnetic fields, it seems that this is the most interesting property of the solutions constructed here that the influence of initially presenting  magnetic fields (even enough strong ones) can be almost negligible at the very beginning of the evolution of such universe, what allows the initial geometry of the universe to be homogeneous and isotropic, but this influence can become much more significant at later stages of its evolution.

\bigskip
\subsection*{Acknowledgments}
The author is grateful to A.N.Golubiatnikov for useful comments.
The author express his deep thanks to ICRANet (Pescara, Italy) and to the Institut des Hautes Etudes Scientifiques (Bures-sur-Yvette, France) for financial support and hospitality during his visits in June and in October 2012 respectively, when the parts of this work were made. The work of the author was also supported in parts by the Russian Foundation for Basic Research (grants 11-01-0034, 11-01-00440) and the program "Fundamental problems of Nonlinear Dynamics" of Russian Academy of Sciences.

\end{document}